# Data-Driven Evolution of Library and Information Science Research Methods (1990–2022): A Perspective Based on Fine-grained Method Entities

Chengzhi Zhang*, Yi Mao, Shuyu Peng
Department of Information Management, Nanjing University of Science and Technology, Nanjing, 210094, China

**Abstract**: Since the 1990s, advancements in big data and information technology have increasingly driven data-centric research in the field of Library and Information Science (LIS). To assess the influence of this data-driven research paradigm on the LIS discipline, this study conducts a fine-grained analysis to uncover the evolutionary trends of research methods within the domain. Using academic papers from LIS published between 1990 and 2022, four key categories of data-driven method entities are automatically extracted: algorithms and models, data resources, software and tools, and metrics. Based on these entities, the study examines the evolution of LIS research methods from three dimensions: the characteristics of research method entities over time, their evolution within different research topics, and the evolutionary features of research method entities across various research methods. The findings highlight data resources as a pivotal driver of methodological evolution in LIS, revealing a cyclical pattern of "emergence-stability/practical application" in the development of research methods within the field.

## 1. Introduction

The *Wechsler Dictionary* defines scientific research methods as "the principles and procedures for the systematic pursuit of knowledge, including the identification and formulation of questions, the collection of data through observation and experimentation, and the formulation and testing of hypotheses" (Lou et al., 2021). In the context of rapid advancements in big data and information technology, scientific research methods are gradually shifting towards data-driven approaches, leading to profound transformations in research practices. Scholars can utilize a wider and more diverse array of data, particularly as data-driven research methods facilitate deeper exploration of problems and refinement of theories (Tang et al., 2021a). Studying how research methods have evolved helps scholars look at problems in new ways and better understand the patterns of how a discipline develops.

* Corresponding author: Chengzhi Zhang (zhangcz@njust.edu.cn).

Since the 1990s, technologies such as the World Wide Web, online social media, big data, and artificial intelligence have rapidly applied to the field of Library and Information Science (LIS). Consequently, scholars have increasingly adopted data-driven research methods, as evidenced by the extensive application, citation, and discussion of related entities in LIS academic papers. It has been suggested that the LIS field is transitioning from empirically driven qualitative research to data-driven quantitative research (Järvelin & Vakkari, 2021). Currently, scholars primarily employ a bibliometric perspective to explore the evolution of theory applications or research methods (Zhang et al., 2023) by examining the textual characteristics of theory and method usage. However, this analysis is coarse-grained and lacks a discussion of the research methods themselves during the evolution process. Investigating research methods from a data-driven perspective can provide deeper insights into how they evolve in response to changes in data resources and technology, offering valuable guidance for future research. Thus, to explore the application of the data-driven research paradigm in the LIS field, this paper extracts four types of research method entities (REMs) related to data-driven research from academic papers published between 1990 and 2022: algorithm and model, data resource, software and tool and metric. Research method entity is a named entity that represent a specific method, for example, nouns or noun phrases used by authors in academic literature for specific means of solving problems (Wang et al., 2022). A fine-grained method entity is a research method entity that has been finely categorized. By analyzing these method entities using objective quantitative data, this study aims to aims to enhance the interpretability of the evolution of research methods in LIS.

Overall, this study extracts four types of fine-grained research method entities from academic papers to capture the evolutionary characteristics of research methods in the field of LIS based on the evolution of research method entities across different periods, research topics, and research methods.

This study explores the evolution of research methods based on fine-grained method entities as an example in the field of LIS. By integrating academic paper topics and research methods, it measures and analyzes the evolutionary characteristics of research methods in the LIS field. Firstly, it is necessary to construct a full-text corpus of academic papers in the LIS field. Next, the structural elements of the academic papers in the corpus need to be identified, because we believe that the methods in the chapter of methods are more likely to be the methods used in the academic papers, so we need to identify the methods chapter of academic papers. Subsequently, the RMEs mentioned in the methods chapter of academic papers should be extracted. Finally, based on the distribution differences of the method entities, the evolutionary characteristics of research methods in the field can be revealed, and an in-depth analysis can be conducted from the perspective of knowledge entities.

To analyze the evolution of research methods from the perspective of method entities, this study assesses the varying distribution of four types of method entities across various years. This analysis is enhanced by examining the distribution of method entities under different topics and methods. By doing so, the study aims to reveal the distinctive features of researchers' utilization of research methods at different time periods, thus delving into a comprehensive exploration of the evolutionary characteristics of research methods based on fine-grained method entities. Based on the above research objectives, the study examines the following four research questions:

***RQ1***: Can the research methods outlined in the Methods chapter serve as a vehicle for the evolution of research methods?

***RQ2***: How has the use of research methods evolved over time through the lens of research method entities?

***RQ3***: In what ways do different research topics shape the evolutionary patterns of research methods?

***RQ4***: What are the evolutionary characteristics of method entities under different 把 varying research approaches？

## 2. Literature review

### *2.1 Evolution of research methods in LIS*

The evolution of research methods has always been a hot research topic in the LIS field, and numerous scholars have conducted extensive research on this subject. Lou et al. (2021) believed that there is an evolution of research methods in LIS. They manually annotated 2,421 papers in the LIS field and found that research methods in LIS underwent a change from focusing on theoretical research to adopting a mixed research approach between 1991 and 2005. Similarly, Järvelin & Vakkari (2021) provided a comprehensive overview of the evolution of research methods in the LIS field over the past 50 years. The analysis revealed a growing popularity of empirical research methods, with survey methods, scientometric methods, experimental methods, and case study methods being employed more frequently. Further support for this trend was found in Vakkari et al. (2022) examination of research development in the LIS field over the past fifty years. It indicated a gradual replacement of conceptual strategies by empirical research as the primary research approach in LIS. Additionally, scientific and professional communication emerged as a major research topic in the LIS field. Furthermore, significant research efforts have been dedicated to method categorization (Chu, 2015; Chu & Ke, 2017), research themes (Han, 2020; Zhang et al., 2023), and the study of gender disparities in scholarly research (Nunkoo et al., 2020; Zhang et al., 2023).

### *2.2 Identification of the chapter structure of the literature*

In academic papers, different chapters serve distinct purposes. The methodology section primarily presents the research methods or design employed in the study, with the methods described therein most likely being those actually applied in the paper. Based on this assumption, the current study aims to identify the structural components of academic papers.

Based on the granularity of annotation, classification logic, and diverse research application fields, the functional classification system of academic paper structures can be categorized into three types: linear classification systems represented by the IMRaD model (Burrough-Boenisch, 1999), modular classification systems represented by the Harmsze model (Harmsze, 2000) and the ABCDE model ( de Waard & Tel, 2006), and argument-based models primarily represented by pragma-dialectics.

The IMRaD model, utilizing sections as the smallest level of annotation granularity, classifies the functional structure of academic paper sections into four main parts: Introduction, Method, Results, and Discussion. The Method section primarily explains the specific research methods or research design employed to address the research problem (Burrough-Boenisch, 1999). The Harmsze model and the ABCDE model categorize sections based on the functions of the literature, with the smallest annotation granularity being at the sentence level. Waard (2007) proposed the pragma-dialectic model based on the ABCDE model, which includes five main categories: facts, research problem, research objectives, method, structure, insights, and hypotheses. In addition, rule-based matching and machine learning methods can be used to automatically identify the structural components of articles.

### *2.3 Extraction of method entities*

The entities of research methods represent the methodological elements in academic papers, providing specific information about the use of research methods. Extracting and studying these method entities from scientific literature is instrumental in understanding the current development status of research methods (Wang et al., 2022) and comprehending the evolutionary process of research methods.

Ibrahim (2021) conducted the identification and manual annotation of statistical software used by scholars in the LIS field to explore the characteristics of their usage in scientific research. Traditional machine learning models such as the hidden Markov model (Rabiner, 1989), maximum entropy Markov model (McCallum et al., 2000), conditional random fields (Lafferty et al., 2001), and support vector machines (Cortes & Vapnik, 1995) have been widely applied in tasks related to method entity extraction, demonstrating excellent performance. The advent of pre-trained language models in the field of natural language processing has revitalized the task of entity extraction (Zhang et al., 2024). Moreover, the recognition of named entities in the biomedical field has also yielded promising results (Duck et al., 2015; Kalim & Mercer, 2022).

# 3. Data and methodology

The objective of this study is to examine the evolution of research methods in the LIS field by identifying the structural components of academic papers and extracting fine-grained method entities. It aims to explore the distribution differences of method entities in academic papers from 1990 to 2022. The framework of this study is illustrated in Figure 1.

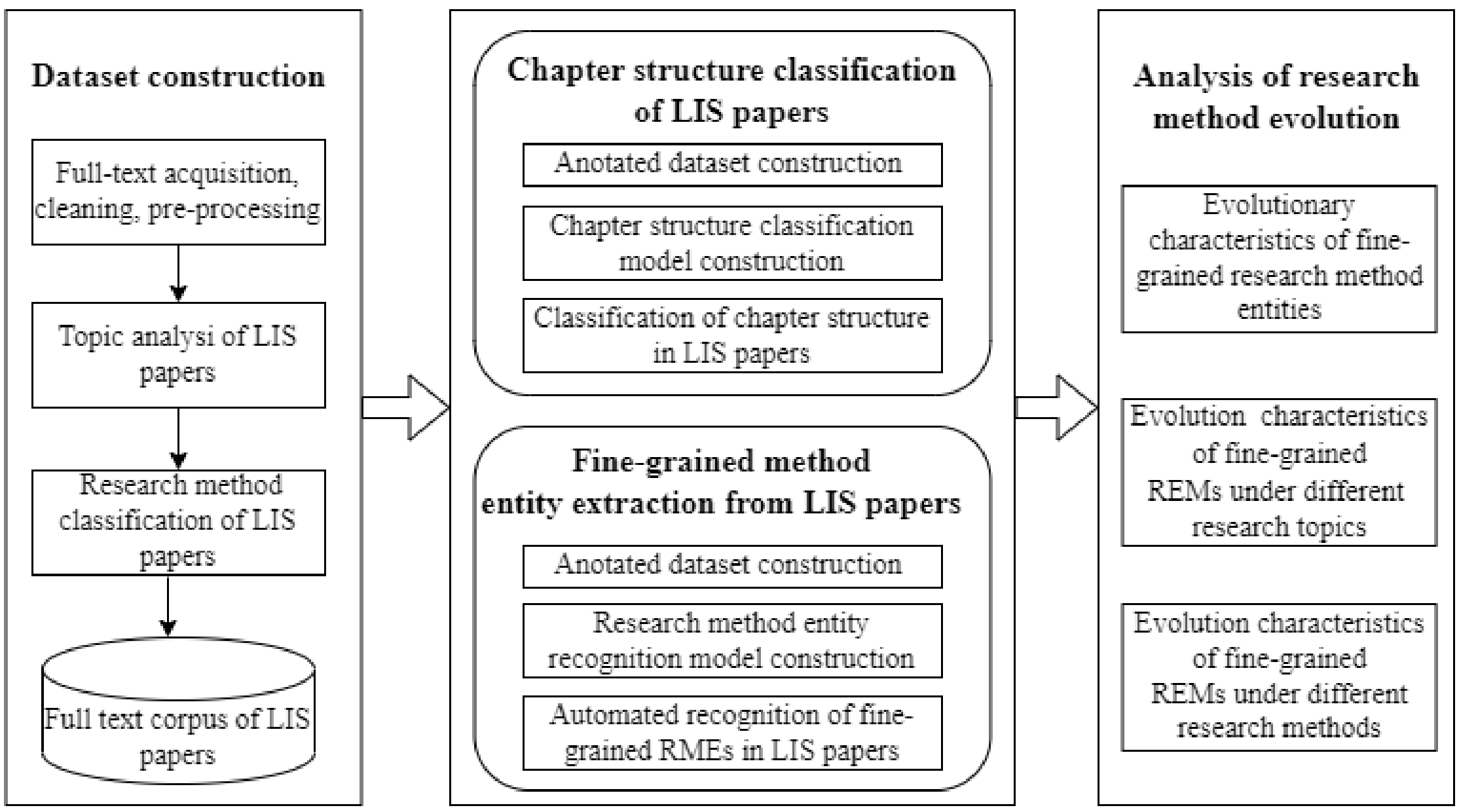

**Figure 1. Framework of this study.**

This section begins by introducing the key steps of the framework. Sub-section 3.2 classifies the research methods employed in LIS papers, while sub-section 3.3 addresses the classification of chapter structures. Sub-section 3.4 details the extraction of fine-grained method entities. Additionally, methods for measuring the chapter-wise distribution of these entities (sub-section 3.5) and analyzing their distributional differences (sub-section 3.6) are presented.

### *3.1 Data Collection*

By combining the representative academic journals in the field identified by Järvelin and Vakkari (2021) and top three quartiles(Q1~Q3) academic journals listed in the JCR2021 (Journal Citation Reports) core journal list, a final list of 14 high-quality representative journals in the field of LIS has been compiled and presented in Table 1. The data collection included metadata and full-text information spanning from 1990 to 2022. A total of 26,100 records were initially collected, including 569 noisy data entries and 163 non-academic documents. After filtering, 25,368 valid data entries remained. Due to the nature of full-text quantitative analysis, 17 papers with missing abstracts and full-text content were excluded from the analysis, resulting in a final sample of 25,351 articles.

**Table 1. Number of articles in the journal.**

| Journal | #Article |
|---|---|
| *Journal of the Association for Information Science and Technology* | 3813 |
| *Journal of Documentation* | 1531 |
| *Library and Information Science Research* | 736 |
| *International Journal of Information Management* | 1784 |
| *Information Processing and Management* | 2707 |
| *Scientometrics* | 5620 |
| *Online Information Review* | 1591 |
| *Aslib Journal of Information Management* | 1299 |
| *The Electronic Library* | 1812 |
| *Journal of Information Science* | 1404 |
| *Journal of Librarianship and Information Science* | 803 |
| *College and Research Libraries* | 1280 |
| *Information Technology and Libraries* | 558 |
| *The Library Quarterly* | 593 |
| **Total** | **25531** |

### *3.2 Research methods classification of LIS papers*

To examine the distribution of research method entities in academic papers employing different research methods, it is essential to categorize the research methods utilized in these papers. Chu and Ke (2017) encoded academic articles from three core journals in LIS—*Journal of the Association for Information Science and Technology (JASIST), Library and Information Science Research (LISR) and Journal of Documentation (JOD)*. They proposed a classification system that includes 16 research methods based on data collection. These methods are listed in Appendix A. Zhang and Tian (2023) applied this classification system to analyze a total of 2,799 academic papers, including those from the aforementioned journals, and developed an automated classification model for research methods based on full-text cognition. In simple terms, this model constructs a summary network that first generates text summaries of sentences related to research method descriptions within the full content of academic papers. Subsequently, it employs these research method summaries for automatic identification and classification using a multi-head attention mechanism network. It is important to note that when utilizing this model for identifying research methods, a single academic paper may be assigned multiple category labels (Zhang & Tian, 2023). The results of the research method identifications are presented in Appendix A.

### *3.3 Chapter structure classification of LIS papers*

Generally speaking, the research methods mentioned in the Method section can represent the specific research methods adopted in the paper, it becomes necessary to first identify the structural components of academic papers in order to locate the Method section before extracting fine-grained method entities. This process lays the foundation for subsequently extracting relevant entities from the "Methods" section, thereby enhancing the relevance and validity of the extracted research method entities in their practical application within the paper. However, due to the current lack of high-quality annotated datasets on academic paper structural components for this study, a combination of manual annotation and machine annotation was employed to construct a corpus for annotating the structural components of academic papers in the LIS field. Furthermore, a stratified sampling method was used to extract 900 samples from the full-text corpus of academic papers in the LIS field, forming the experimental dataset.

In addition, this study incorporated the literature review functional section into the IMRaD framework (Burrough-Boenisch, 1999) to further enhance the applicability of this classification system in the field. Furthermore, by referring to the chapter structure classification system in the study by Lu et al. (2018), this research ultimately established a six-category structural classification system for academic papers in the LIS field, including Introduction, Related Work, Method, Evaluation and Results, Discussion and Conclusion, and Other.

This study selected traditional machine learning models, deep learning models based on pre-trained language models, and large-scale language models based on T5 for experimentation, testing the performance of a total of eleven specific models across these three categories. For the chapter structure categorization task, three traditional machine learning algorithms were employed, including support vector machines, random forests, and logistic regression. Word embedding was performed using the BERT model, which is based on pre-trained language models, and four different deep neural networks were experimented with. Additionally, three versions of the T5 large model (Raffel et al., 2020) were also used for structural classification experiments: T5_large, T5_base, and T5_small.

Table 2 presents the performance metrics obtained from training models using chapter titles and content features. Among these eleven models, the large language model performed the best in structural classification, with the T5_base model achieving the most effective structural classification, attaining a weighted average $F_1$ score of 84.23%.

Consequently, this study employed the T5_base model trained on chapter titles and content to identify the structural components of academic papers in the corpus. Among them, articles with ambiguous sections were not subjected to structural identification. Ultimately, structural function prediction was performed on 23,012 academic papers, encompassing a total of 153,230 sections, as shown in Appendix B.

**Table 2. Performance comparison of chapter structure classification models based on "Chapter Title + Chapter Content" features.**

| Model | | Macro Avg. | | | Weight Avg. | | |
|---|---|---|---|---|---|---|---|
| | | **P(%)** | **R(%)** | **$F_1$(%)** | **P(%)** | **R(%)** | **$F_1$(%)** |
| Traditional ML | LR | 84.44 | 82.11 | 82.95 | 84.61 | 83.62 | 83.81 |
| | SVM | 82.05 | 80.79 | 81.17 | 83.09 | 82.41 | 82.58 |
| | RF | 79.26 | 73.29 | 75.02 | 77.21 | 75.86 | 75.41 |
| DL | SciBERT | 81.36 | 79.43 | 80.18 | 82.27 | 81.80 | 81.85 |
| | SciBERT_CNN | 80.30 | 80.40 | 80.29 | 82.41 | 81.80 | 82.05 |
| | SciBERT_DPCNN | 81.69 | 79.18 | 80.14 | 81.58 | 81.25 | 81.20 |
| | SciBERT_RCNN | 79.93 | 79.90 | 79.78 | 81.77 | 81.42 | 81.46 |
| | SciBERT_BilSTM | 80.90 | 80.88 | 80.69 | 82.76 | 81.80 | 82.07 |
| LLM | T5_small | 83.45 | 81.78 | 82.17 | 84.53 | 83.97 | 83.84 |
| | T5_base | 83.61 | 82.78 | 82.94 | 85.00 | 83.97 | 84.23 |
| | T5_large | 79.92 | 79.40 | 79.44 | 82.17 | 81.83 | 81.55 |

### *3.4 Fine-grained method entity extraction from LIS papers*

In this study, an annotated corpus of fine-grained method entities in LIS was initially constructed. Drawing on relevant studies (Wang et al., 2022), algorithm and model entities, data resource entities, software and tool entities, and metrics entities were extracted from the research papers. These entities typically encapsulate key information within the papers, aiding researchers in comprehending tasks, task evaluation, and methodological innovation more effectively. The definitions of the various types of research method entities (RMEs) are shown in Appendix C. It is important to note that in this study, RMEs constitute the overarching class concepts encompassing the four types of fine-grained entities listed in the table in Appendix C, all of which fall under the category of research method class entities.

We constructed a silver standard annotated corpus for fine-grained entity extraction. First, we utilized ChatGPT to generate fine-grained research method entity terms in the LIS field. After manual verification and supplementation, we compiled a list containing these four categories of research method entities. Finally, this list was matched against the method entities in 27,637 Methods sections to obtain the silver-standard annotated corpus for research method entities, thereby preparing for subsequent model training. The silver standard annotated corpus underwent sampling and manual verification to enhance the accuracy and recall of the annotations. Based on syntactic structure, this corpus was segmented into sentences, resulting in a total of 2,742 sentences containing four types of fine-grained method entities, including 588 algorithm and model entities, 481 data resource entities, 354 software and tool entities, and 438 metric entities.

Building upon previous research, we adopted the BERT+BiLSTM+CRF model architecture (Ma & Yuan, 2019) to construct a fine-grained method entity recognition model. This approach involved leveraging BERT (Devlin et al., 2019) for textual semantic encoding and embedding, utilizing BiLSTM network (Graves & Schmidhuber, 2005) for semantic learning with the context of academic text, and applying CRF (McCallum & Li, 2003) for sequence labeling. Furthermore, during the text embedding layer testing phase, both BERT_base and SciBERT versions of the BERT model were utilized. Additionally, both the BERT+CRF and BERT+SOFTMAX models were used as baseline models for comparative experiments on the original annotated data.

The experimental results are shown in Table 3, indicate that the SciBERT+BiLSTM+CRF model achieved the best overall performance, with precision, recall, and $F_1$ values of 71.43%, 75.42%, and 73.38%, respectively, on the task of recognizing fine-grained research method entities. Specifically, the model exhibited the best identification performance for algorithm and model entities, with an $F_1$ score of 78.45%. This may be attributed to the fact that algorithm typically have more well-defined and standardized terminologies, making them easier to identify accurately. The $F_1$ scores for data resources, software and tools, and evaluation metrics were 73.61%, 66.36%, and 70.19%, respectively.

**Table 3. Performance comparison of fine-grained research method entity recognition models.**

| Model | Performance indicators (%) | | | | | | | | | | | | | | |
|---|---|---|---|---|---|---|---|---|---|---|---|---|---|---|---|
| | Algorithm & Model | | | Data resource | | | Soft & Tool | | | Metric | | | Comprehensive | | |
| | P | R | $F_1$ | P | R | $F_1$ | P | R | $F_1$ | P | R | $F_1$ | P | R | $F_1$ |
| BERT+CRF | 54.84 | 75.98 | 63.70 | 55.36 | 66.67 | 60.79 | 50.00 | 52.31 | 51.13 | 55.96 | 46.21 | 50.62 | 54.56 | 62.47 | 58.25 |
| BERT+Softmax | 62.56 | 76.54 | 68.84 | 60.00 | 64.52 | 62.18 | 47.06 | 49.23 | 48.12 | 60.47 | 59.09 | 59.77 | 59.50 | 65.46 | 62.34 |
| BERT+BiLSTM+CRF | 63.40 | 73.04 | 67.88 | 71.11 | 71.11 | 71.11 | 54.44 | 51.04 | 52.69 | 55.47 | 69.61 | 61.74 | 62.07 | 67.97 | 64.89 |
| SciBERT+CRF | 70.10 | 79.89 | 74.67 | 60.00 | 61.29 | 60.64 | 47.19 | 64.62 | 54.55 | 58.11 | 65.15 | 61.43 | 61.19 | 69.94 | 65.27 |
| SciBERT+Softmax | 72.77 | 82.12 | 77.17 | 63.27 | 66.67 | 64.92 | 50.00 | 67.69 | 57.52 | 53.47 | 58.33 | 55.80 | 62.03 | 70.36 | 65.93 |
| SciBERT+BiLSTM+CRF | 77.51 | 79.41 | 78.45 | 73.88 | 73.33 | 73.61 | 60.18 | 73.96 | 66.36 | 68.87 | 71.57 | 70.19 | 71.43 | 75.42 | 73.38 |

To enhance the performance of the SciBERT+BiLSTM+CRF model, we also conducted data augmentation experiments. Table 4 presents the performance data with varying scales of added data, where ORI denotes the original training set and AUG indicates the augmented data. The percentage represents the proportion of augmented data relative to the original training set, for example, ORI+AUG (5%) means the training data consists of the original training data plus an augmentation of 5% of the original training data. The experimental results demonstrate that data augmentation can improve model performance, with the best-performing SciBERT+BiLSTM+CRF model being trained using the original annotated data combined with 10% of augmented data.

**Table 4. Comparison of optimal model performance with different augmented data.**

| Model / Augmented Data | Performance indicators (%) | | | | | | | | | | | | | | |
|---|---|---|---|---|---|---|---|---|---|---|---|---|---|---|---|
| | Algorithm & Model | | | Data resource | | | Soft & Tool | | | Metric | | | Comprehensive | | |
| | P | R | $F_1$ | P | R | $F_1$ | P | R | $F_1$ | P | R | $F_1$ | P | R | $F_1$ |
| **SciBERT+BiLSTM+CRF** | | | | | | | | | | | | | | | |
| ORI | 77.51 | 79.41 | 78.45 | 73.88 | 73.33 | 73.61 | 60.18 | 73.96 | 66.36 | 68.87 | 71.57 | 70.19 | 71.43 | 75.42 | 73.38 |
| ORI+AUG（5%） | 77.57 | 81.37 | 79.43 | 76.92 | 74.07 | **75.47** | 56.15 | 76.04 | 64.60 | 67.86 | 74.51 | **71.03** | 70.82 | 77.28 | **73.91** ↑ |
| ORI+AUG（10%） | 77.83 | 80.88 | 79.33 | 74.45 | 75.56 | 75.00 | 59.17 | 73.96 | 65.74 | 69.16 | 72.55 | 70.81 | 75.53 | 76.72 | **74.03** ↑ |
| ORI+AUG（15%） | 75.58 | 80.39 | 77.91 | 72.44 | 68.15 | 70.23 | 55.00 | 68.75 | 61.11 | 61.11 | 64.71 | 62.86 | 67.83 | 72.25 | 69.97 |
| ORI+AUG（20%） | 76.78 | 79.41 | 78.07 | 75.40 | 70.37 | 72.80 | 58.47 | 71.88 | 64.49 | 62.83 | 69.61 | 66.05 | 69.89 | 73.93 | 71.86 |
| ORI+AUG（25%） | 79.25 | 82.35 | **80.77** | 73.64 | 70.37 | 71.97 | 56.00 | 72.92 | 63.35 | 65.45 | 70.59 | 67.92 | 70.31 | 75.42 | 72.78 |
| ORI+AUG（30%） | 79.71 | 80.88 | 80.29 | 75.19 | 74.07 | 74.63 | 60.50 | 75.00 | **66.98** | 64.55 | 69.61 | 66.98 | 71.70 | 75.98 | **73.78** ↑ |

Consequently, the tuned model was ultimately employed for the automatic recognition of fine-grained method entities across the entire dataset. Following preprocessing of the 153,230 chapters, a total of 6,145,694 sentences were obtained as the model input for predicting fine-grained method entities, as illustrated in Appendix D.

### *3.5 Measurement of chapter distribution of method entities*

To determine the chapter distribution characteristics of research method entities, we measured the average distribution density of these entities within chapters. The chapter distribution density $\rho$ of RMEs is defined as the ratio of the total number of tokens for all research method terms within a chapter to the overall number of tokens in that chapter. The average chapter distribution density $\bar{\rho}$ of RMEs is the arithmetic mean of the chapter distribution densities of RMEs in chapters of the same type. The formulas for calculating the chapter distribution density $\rho$ and the average chapter distribution density $\bar{\rho}$ of RMEs are presented as Equations 1 and

2.

$$\rho_i = \frac{\sum_x^l n_x}{N_i} \quad (1)$$

$$\bar{\rho}_c = \frac{\sum_j^{count_c} \rho_j}{Count_c} \quad (2)$$

Where $\rho_i$ represents the distribution density of research method entities in chapter $i$, $N_i$ denotes the total number of tokens in chapter $i$, and $n_x$ indicates the number of tokens for the $x$-th research method entity in chapter $i$. Additionally, $\bar{\rho}_c$ is the average distribution density of research method entities in chapters of category $c$, and $Count_c$ signifies the total number of chapters in category $c$. Before calculating the average distribution density $\bar{\rho}$ of research method entities, we utilized the BERT Tokenizer tool to tokenize both the original text of the chapters and the text of the research method entity terms, transforming them into token sequences.

***3.6 Measurement of the distribution differences of method entities***

This study utilized the term features of fine-grained method entities to evaluate the distribution differences of four types of RMEs across different years, with the aim of exploring the evolution of research methods. To achieve this, the Jaccard distance (Jaccard, 2010) was employed to calculate the distance between the four RMEs groups within two adjacent time windows separately. This helped capture the differences in the distribution of the four RME groups through the Jaccard distance, thus reflecting the variability in the specific usage characteristics of the research methods. This paper defines the Jaccard distance, denoted as $D_J(A, B)$, a measure of dissimilarity between two sets, $A$ and $B$, which contain $m$ and $n$ elements, respectively. The Jaccard distance ranges from 0 to 1, where a higher value indicates greater dissimilarity between the sets. The formula for computing the Jaccard distance is given in Equation 3.

$$D_J(A, B) = 1 - J(A, B) = 1 - \frac{|A \cap B|}{|A \cup B|} \quad (3)$$

In this study, the Jaccard distance is used to quantify the distribution discrepancy of RME groups across different time periods. For example, consider the distributions for one year $Y_t$ and the subsequent year $Y_{t+1}$, defined as:

$RME_t = \{e1, e2, e3, e4, e5\}$,

$RME_{t+1} = \{e1, e3, e4, e6, e7\}$.

Using equation (3), the Jaccard distance between $Y_t$ and $Y_{t+1}$ is calculated as:

$$D_J(RME_t, RME_{t+1}) = 1 - \frac{|RME_t \cap RME_{t+1}|}{|RME_t \cup RME_{t+1}|} = 1 - \frac{|\{e1,e2,e3,e4,e5\} \cap \{e1,e3,e4,e6,e7\}|}{|\{e1,e2,e3,e4,e5\} \cup \{e1,e3,e4,e6,e7\}|} = 1 - \frac{3}{7} = \frac{4}{7}$$

Thus, the distribution discrepancy of RME groups between $Y_t$ and $Y_{t+1}$ is 4/7.

Since the Jaccard distance only captures the disparity in the distribution of RMEs between adjacent time windows, this paper further computed the growth rate of the Jaccard distance to illustrate evolving trends in the specific usage characteristics of research methods. The formula for calculating the growth rate $G$ of the Jaccard distance is presented as Equation 4.

$$G = \frac{D_J(B, C) - D_J(A, B)}{D_J(A, B)} \qquad (4)$$

Where, set $A$, $B$, and $C$ represent the distributions within three consecutive time windows. $D_J(A, B)$ denotes the Jaccard distance between the first two time windows, while $D_J(B, C)$ represents the Jaccard distance between the latter two time windows. When $G$ is positive, its value is positively correlated with the rate of expansion of the Jaccard distance. Conversely, when $G$ is negative, its value is negatively correlated with the rate of reduction of the Jaccard distance.

## 4. Results

Based on the recognition results of chapter structure and the extraction of RMEs, this study computed the Jaccard distance of method entities within different time windows. This analysis serves as a foundation for exploring the evolution of research methods in the LIS domain. Fine-grained method entities can encapsulate specific usage information of research methods, making them a valuable resource for examining the distribution variances and evolutionary trends of research methods in the field.

This section is divided into four sections, each corresponding to one of the four research questions.

### *4.1 Distribution of RMEs in different chapters*

We answer *RQ1* in this section. Using the trained fine-grained automatic research method entity recognition model, all academic papers in the corpus were recognized at the chapter level to identify fine-grained method entities. Figure 2 visualizes the calculation results of the chapter distribution density and the average chapter distribution density of RMEs. Subfigure (a) displays a box plot of the chapter distribution density of RMEs, while subfigure (b) shows a histogram of the average distribution density of RMEs. As depicted in the figure, the average distribution density of RMEs across the five different types of chapters' ranges from 1% to 2%, with the highest average distribution density observed in the Methods chapter at approximately 1.936%. The lowest average density of distribution of RMEs is found in the Related work chapter, at about 1.426%.

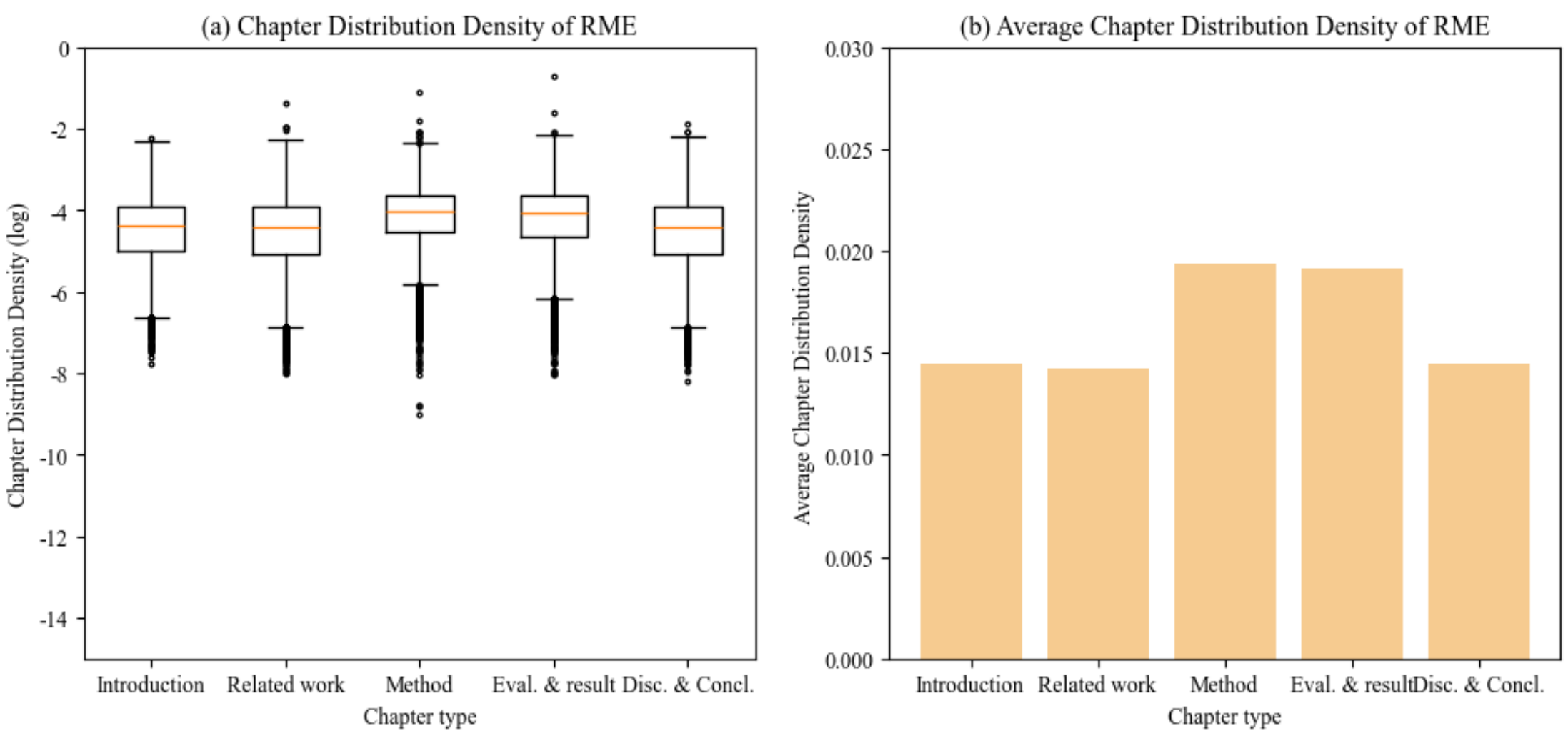

**Figure 2. Chart of fine-grained research method entity chapter distribution.**

Given that the average distribution density of RMEs is highest in the Methods chapters, it can be inferred that the RMEs appearing in these chapters are more likely to represent the actual research methods used in academic papers compared to those in other chapters. Therefore, this study regarded the research methods found in the Methods chapters as carriers for investigating the evolution of research methods. In exploring the evolution of research methods, this study specifically focused on the research methods appearing in the Methods chapters.

### *4.2 Evolutionary characteristics of fine-grained research method entities*

*4.2.1 The overall evolutionary characteristics of the distribution differences of fine-grained RMEs*

We answer *RQ2* in this section. To gain insights into the evolution of research methods within the LIS domain, we examined the annual proportions of four categories of research method entities from 1990 to 2022, as illustrated in Figure 3. Overall, prior to 2004, the proportions of these four entity categories displayed fluctuations. However, around 2015, the shares of data resource entities and software and tool entities began to stabilize, while the proportion of algorithm and model entities showed a gradual increase. This shift may be attributed to the rising prominence of diverse algorithms associated with large language models, leading scholars to increasingly adopt research methods related to algorithms. In contrast, the proportion of metric entities has decreased, which may suggest that the utilization of evaluation metric methods in the LIS field is reaching a more stable state.

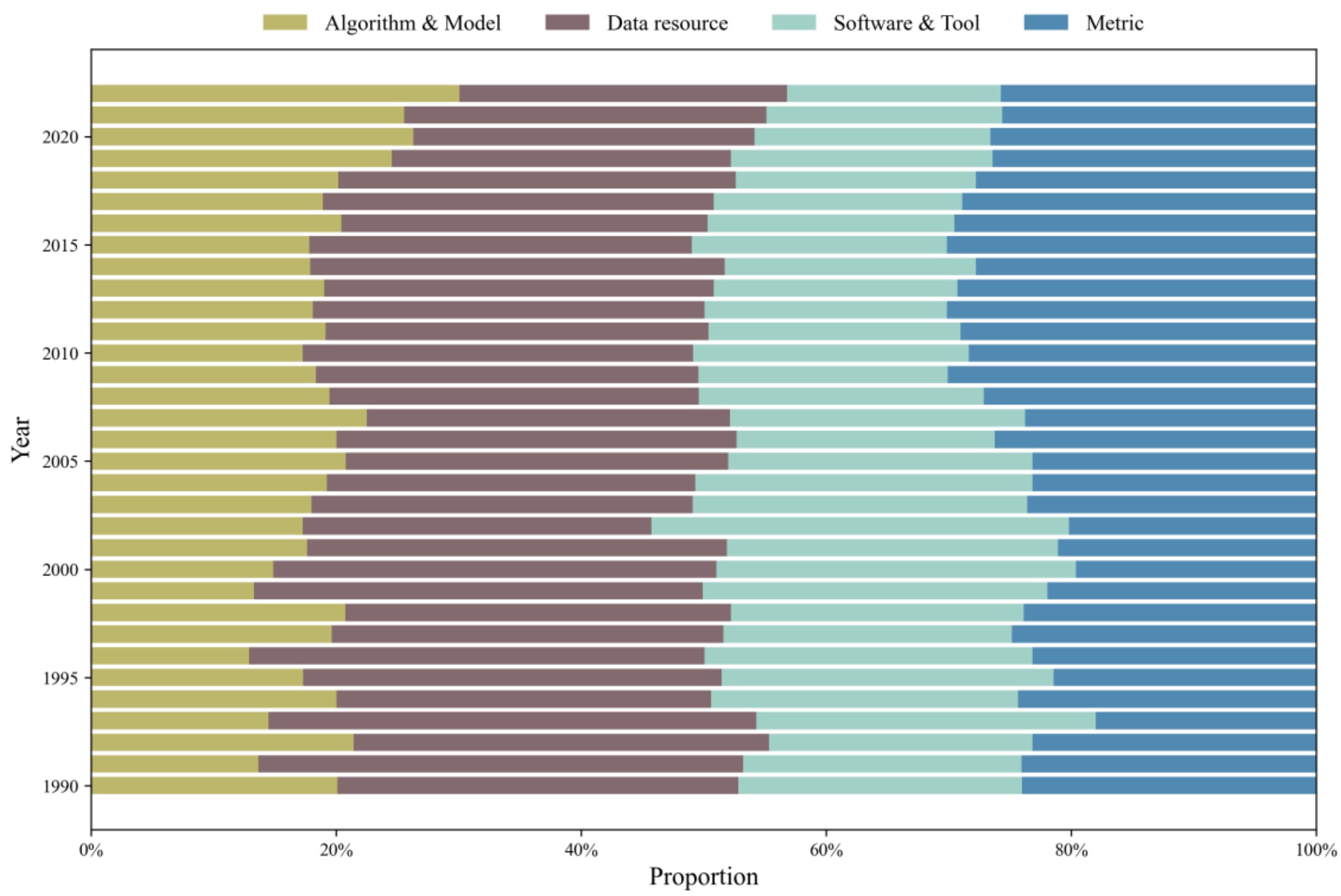

**Figure 3. Distribution chart of proportions of four types of entities across different years.**

We employed the Jaccard distance to measure the distribution differences of research method entities between two adjacent time periods. A smaller Jaccard distance indicates a higher similarity in the use of research methods across the two time windows, while a larger Jaccard distance suggests differences in the usage of research methods within those windows. Our corpus spans years from 1990 to 2022, therefore, we set the calculation time window for the Jaccard distance to 3 years, resulting in a total of 11 time windows and necessitating the computation of 10 pairs of Jaccard distances. The results of these calculations are presented in Figure 4.

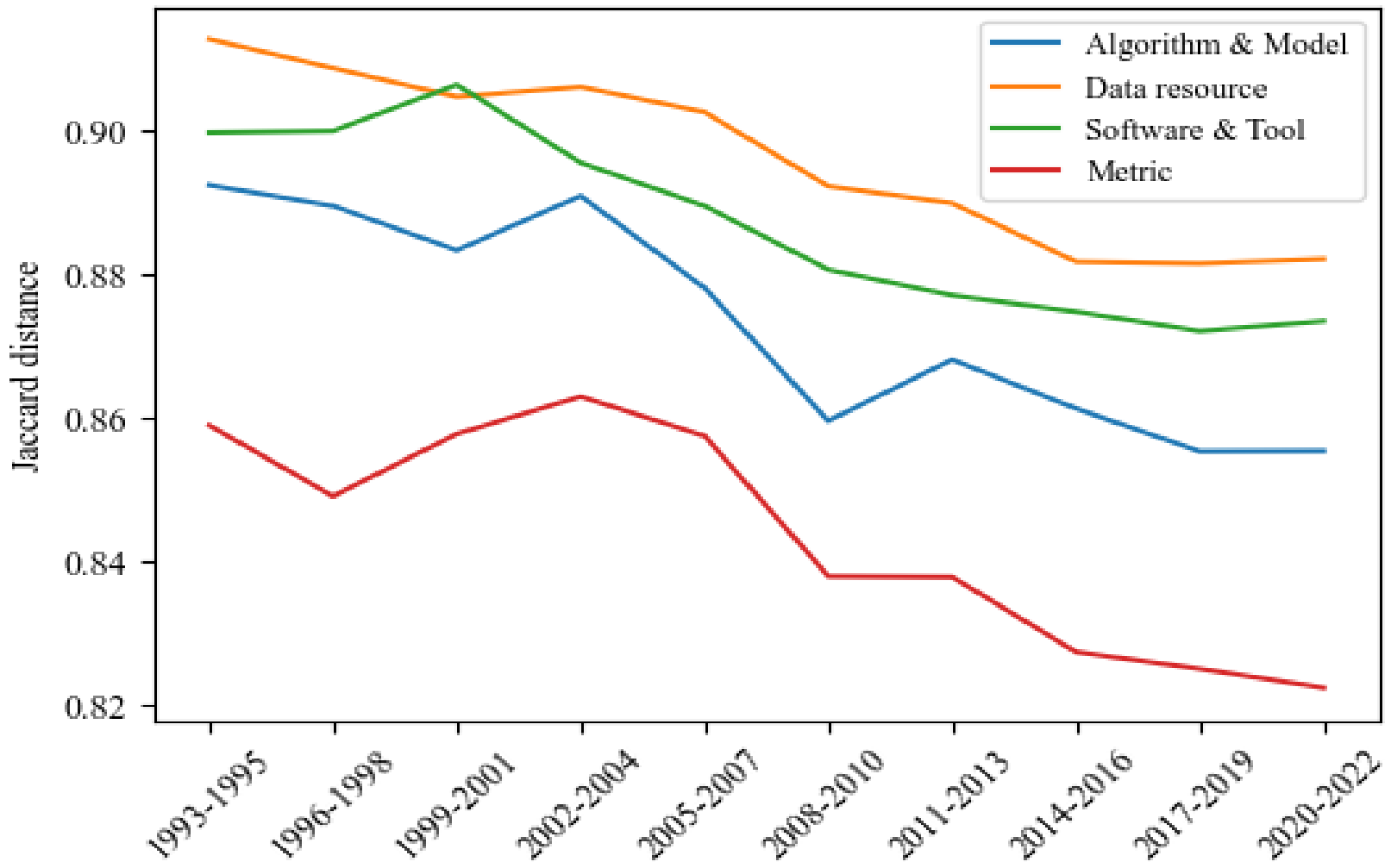

**Figure 4. Evolution of distribution discrepancy of RME groups across different time periods.**

As illustrated in the figure 4, during the majority of periods, the usage differences among the four categories of entities were greatest for data resource entities. This may be attributed to the diversity and complexity of data. In contrast, the usage differences for metrics have consistently been the smallest. This could be due to the fact that effective evaluations often rely on a limited number of universal standards throughout the development of the discipline, leading to a normalization process that results in increasingly similar uses of evaluation metrics. Beginning from the period of 2002-2004, the usage differences for data resources, software and tools, and metrics all showed a trend of decreasing variability, indicating that scholars in the LIS field are gradually stabilizing their selection of these three types of research methods. From 2008 to 2010, the differences in the algorithm and model research method entities exhibited a pattern of first increasing and then decreasing. This fluctuation may have been influenced by the rise of neural networks during this period, which introduced new algorithms and consequently reduced similarity. However, from 2011 to 2013, scholars also began to stabilize their use of algorithm and model entities. Overall, between 1990 and 2022 within the LIS field, the differences in the use of data resource research methods are the largest, suggesting that "data" may serve as a driving force behind academic research, particularly within an increasingly data-driven research landscape.

*4.2.2 The evolving characteristics of changes in the distribution differences of fine-grained RMEs*

We expanded the fluctuations of Jaccard distance curves for different periods of research method entities in the form of Jaccard distance growth rates. Figure 5 presents the calculation results of the Jaccard distance growth rates, where positive values indicate an increase in Jaccard distance, meaning that the differences in the use of research methods are widening, while negative values suggest an enhancement in the similarity of research method usage. From the figure 5, it can be observed that the changes in the differences in research method usage exhibit cyclical fluctuations. In the period from 1996 to 2004, the Jaccard distance growth rates for the four types of research method entities all displayed positive values, indicating that the differences in research method usage characteristics were on the rise. This suggests that during this time, scholars employed more new tools and new data distinct from previous work, marking a phase of "emergence" in research methods. From 2005 to 2010, the differences in research method usage decreased at an increasing rate, with a greater number of previously mentioned research method entities appearing in academic papers. This period can be characterized as a "stabilization" or "practical" phase for research methods. From 2011 to 2016, the aforementioned fluctuation trend re-emerged, differing in that this period was dominated by the emergence of algorithm and model research methods. Between 2017 and 2022, the Jaccard distance growth rates for software and tool entities, algorithm and model entities, and data resource entities approached zero from negative values and exhibited a trend toward positive values. This reflects ongoing innovations in the means of utilizing research methods during this period. Therefore, over the past 30 years in the LIS field, there exists a cyclical process characterized by "emergent periods" and "stabilization/practical periods" in research methods.

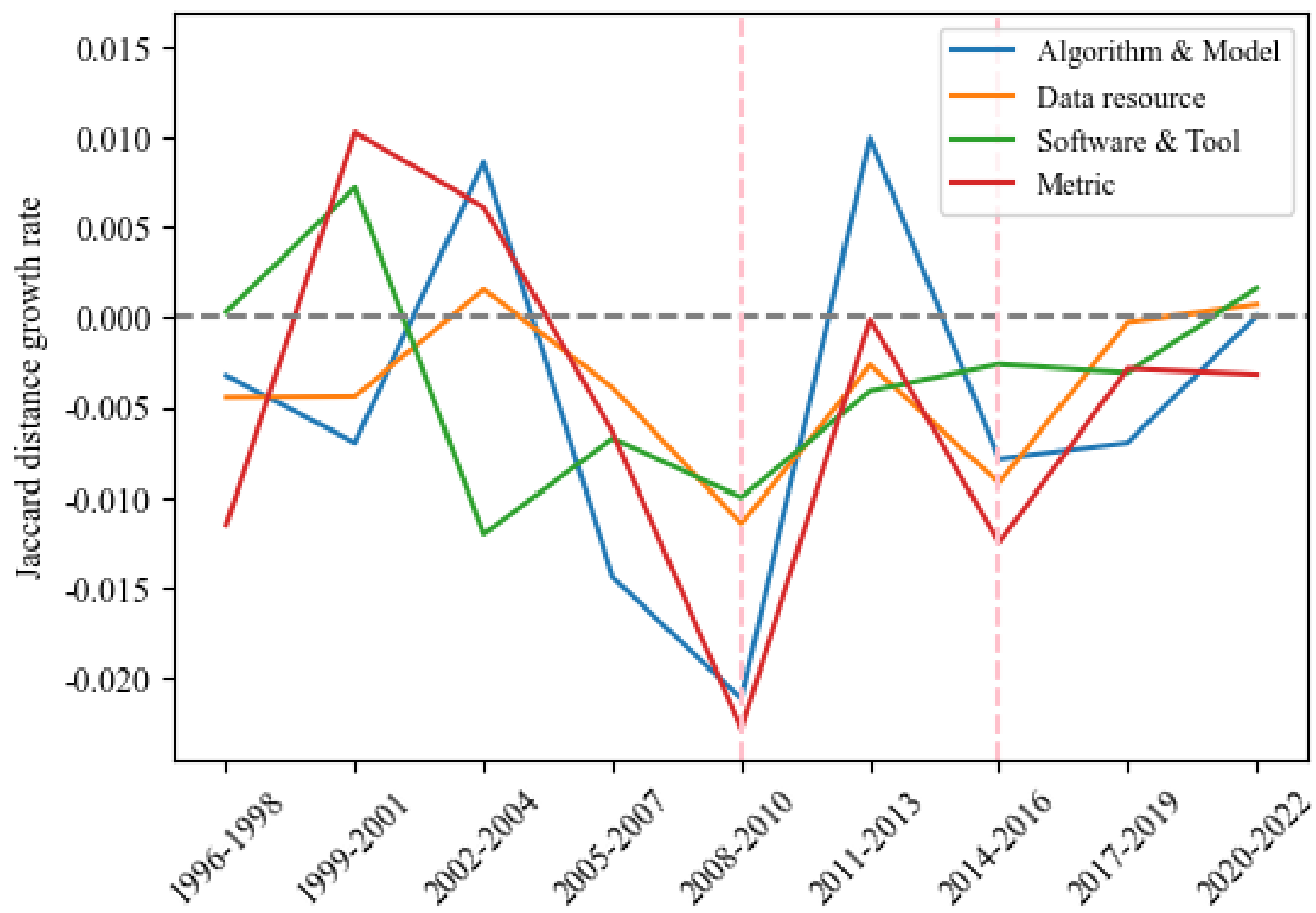

**Figure 5. Evolution of growth rate in distribution discrepancy of RME groups across different time periods.**

### *4.3 Analysis of the evolution characteristics of fine-grained RMEs under different research topics*

We answer *RQ3* in this section. In order to examine the evolutionary characteristics of the distribution variance of fine-grained RMEs in academic papers across different research topics in the field of LIS, we calculated the Jaccard distances of the four categories of RMEs in articles from 1990-1998, 1999-2006, 2007-2014, and 2015-2022. We utilized the Top2Vec model for topic modeling analysis of the corpus (Angelov, 2020) and employed the Coefficient of Variance as the coherence measure to determine the optimal number of topics. The coherence score was maximized when the number of topics was set at 19. Thus, this study determined the number of research topics in the full-text original corpus of academic papers in the field of LIS to be 19. We summarized and named 19 research themes by combining the representative keywords generated by Top2Vec for each topic. Table 5 presents the corresponding names of these research themes. It is important to note that our summary of "Metadata / Cataloging" pertains to traditional libraries and archival management, while "Knowledge Organization" is intended to address the complex information needs in digital environments.

**Table 5. Research topics.**

| Research topic | #Article | |
|---|---|---|
| Information Retrieval | 2008 | 7.92% |
| Informetrics / Scientometrics | 1748 | 6.90% |
| Scientific Evaluation | 1719 | 6.78% |
| Information Ethics / Information Philosophy | 1684 | 6.64% |
| Knowledge Management | 1664 | 6.56% |
| Webometrics / Altmetrics | 1640 | 6.47% |
| Digital Information Resources | 1615 | 6.37% |
| Information Seeking | 1390 | 5.48% |
| User Information Behavior | 1354 | 5.34% |
| College Library Service | 1313 | 5.18% |
| Health Information | 1299 | 5.12% |
| Knowledge Mapping | 1218 | 4.80% |
| Text Ming | 1213 | 4.78% |
| Information Literacy | 1062 | 4.19% |
| Metadata / Cataloging | 1048 | 4.13% |
| Digital Library | 1015 | 4.00% |
| Patent Research | 794 | 3.13% |
| Librarian Profession | 788 | 3.11% |
| Knowledge Organization | 779 | 3.07% |
| Total | 25351 | 100% |

We calculated the Jaccard distances of the four types of research method entities in articles across different research themes. It is important to note that Figure 6 presents the results of the Jaccard distance calculations between two time windows.

The research results reveal that academic papers on various research topics generally exhibit a steady decrease or a stable declining trend in the specificity of research method usage characteristics. This reflects a gradual convergence in the patterns of research method utilization by scholars when investigating specific topics.

Specifically, the distribution differences of algorithm and model entities and software and tool entities across different research topics exhibit various trends, including stable, declining, rising, declining then rising, and rising then declining patterns. In contrast, the distribution differences of data resource entities across different research themes show a declining trend, as well as a pattern of declining followed by rising. Meanwhile, the distribution differences of metric entities across different research themes demonstrate trends of declining, rising, and both declining then rising and rising then declining. Regarding the distribution variance of data resource research method entities, research topics like Information Seeking and Informetrics / Scientometrics also show a trend of initially decreasing and then increasing. This trend may be attributed to the growing influence of the open access movement and the continuous enhancement in both the quantity and quality of online information resources in recent years (Vakkari, 2024). Scholars have been able to utilize a more diverse and abundant range of data for research in these areas, making data an important factor in the evolution of research methods. In the research topic of Scientific Evaluation, besides metric research method entities, the distribution variance of the other three categories of research method entities is decreasing. The distribution variance of metric research method entities shows a trend of initial increase followed by decrease, reflecting a stabilization in the development of evaluation indicators in recent years.

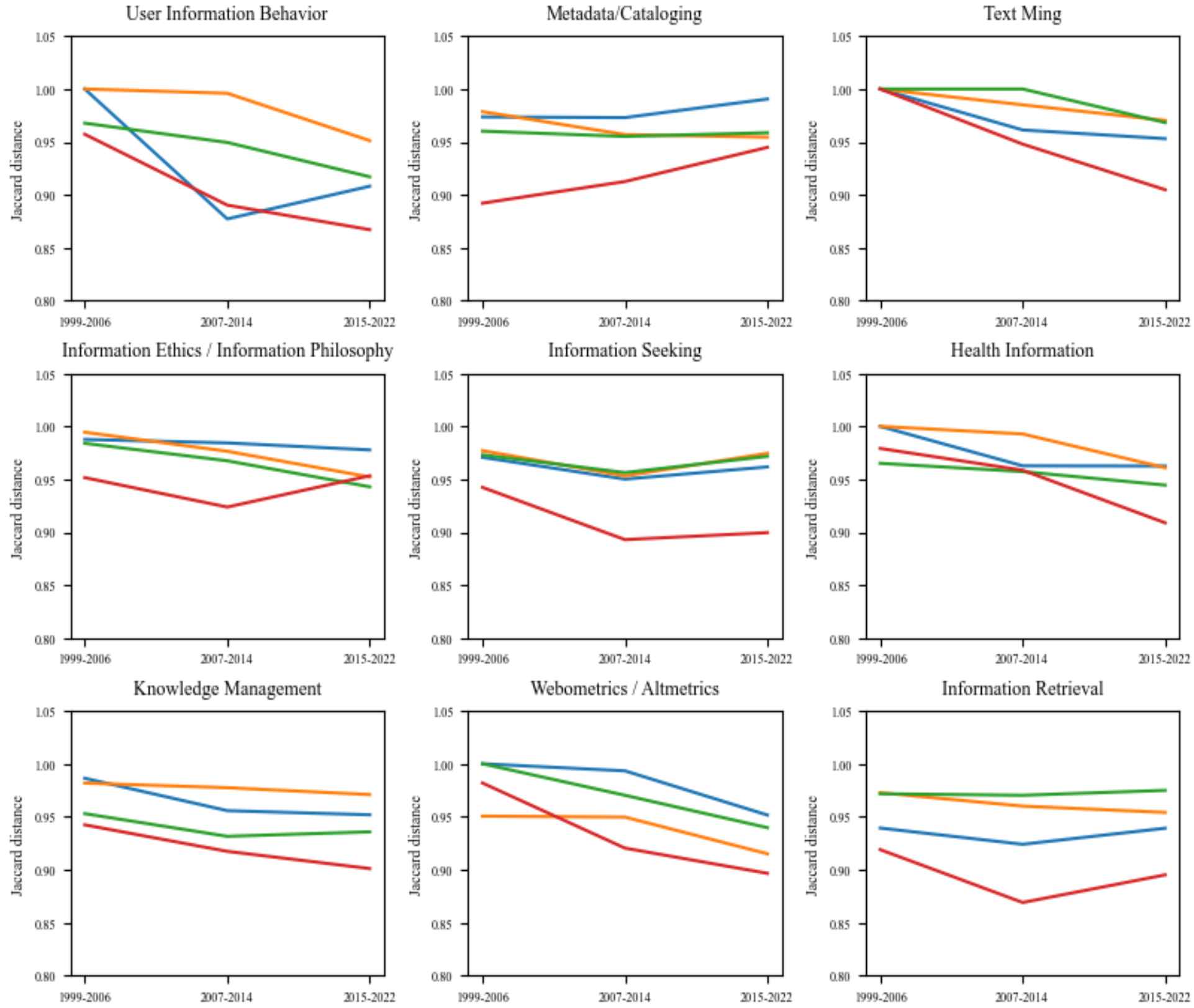

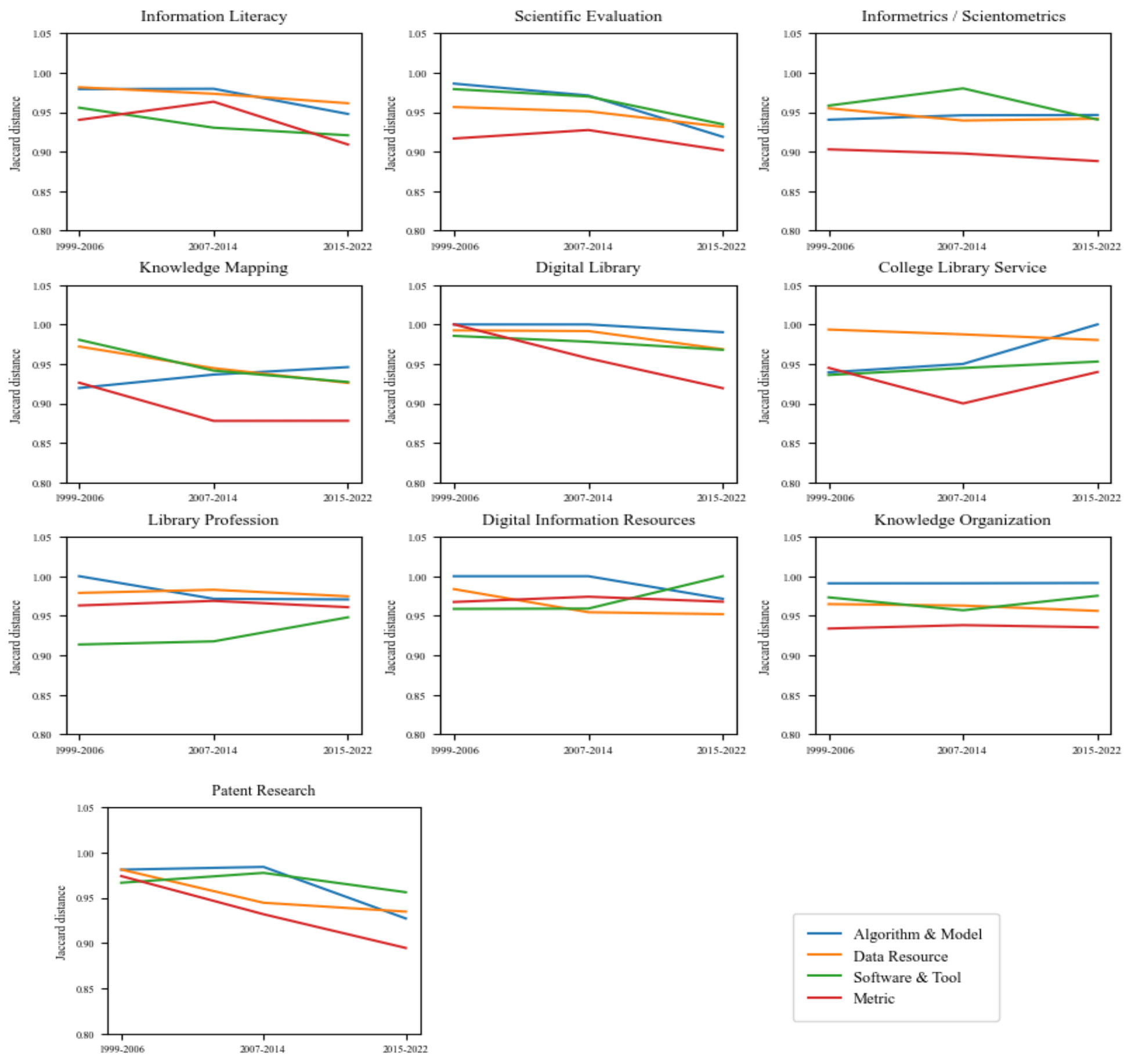

**Figure 6. Evolution of distribution discrepancy of fine-grained RMEs across different research topics.**

### *4.4 Evolutionary characteristics of fine-grained method entities under different research methods*

We answer *RQ4* in this section. We further explored the distribution variance of fine-grained method entities in academic papers across different research methods in the field of LIS. This study employed the research method classification system proposed by Chu and Ke (2017) as the standard for classifying research methods in the corpus of academic papers. Furthermore, an automatic classification model of research methods based on full-text cognition proposed by Zhang and Tian (2023) was adopted to identify research methods at the document level in academic papers, with each academic paper potentially being assigned multiple research method category labels.

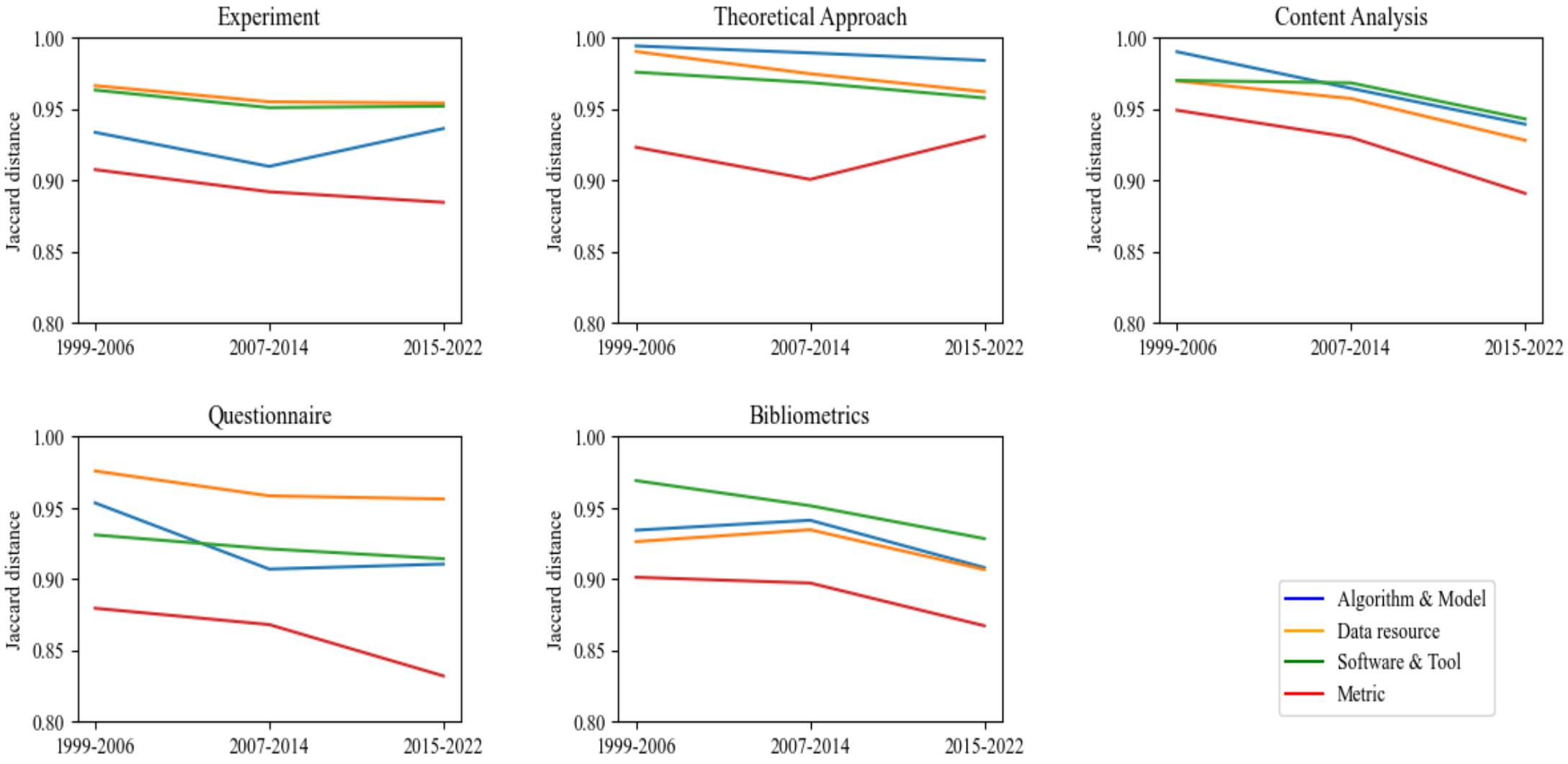

**Figure 7. Evolution of distribution discrepancy of fine-grained RMEs across different research methods.**

Due to the uneven distribution of the quantity and temporal occurrence of academic papers using different research methods, this section only calculated the Jaccard distances of four categories of RMEs in articles on the five most frequently used research methods across four time windows. These five frequently used research methods are experiment, theoretical approach, content analysis, questionnaire, and bibliometrics. Figure 7 illustrates the distribution variance of fine-grained method entities under different research methods. It can be observed that in academic papers using experimental methods, the distribution variances of software and tool entities, data resource entities, and algorithm and model entities all exhibit an early decrease followed by a recent increase, especially with a marked increase in the distribution variance of algorithm and model entities in the most recent time window. The experimental method is a commonly used research approach for data-driven studies in the field of LIS. From the perspective of the distribution variance of RMEs, with the development and progress of information science, the usage patterns and method connotations of the experimental method are continuously evolving. In content analysis and bibliometrics methods, various types of RMEs exhibit a sustained decrease in distribution variance, reflecting, to some extent, the relatively stable patterns of usage associated with these research methods.

## 5. Discussion

### *5.1 Reasons for the evolutionary characterization of research methods*

In the LIS field, previous studies have often discussed the evolution of research methods from a qualitative perspective. However, this study delves into fine-grained method entities to examine the evolution of research methods from a more microscopic perspective. The research reveals the diversity in method usage, innovation in method development, and the evolving patterns driving method development over time in the LIS discipline.

Our study spans from 1990 to 2022. By analyzing the distribution variance of RMEs across different research topics, we discovered that data resources play a breakthrough role in the generation of new research. This finding may stem from the advent of the data intelligence era and the rapid development of artificial intelligence technology, which have led to the continuous emergence of open access and online information resources. These resources provide scholars with richer data sources, thereby driving the breadth and depth of scientific research (Tang et al., 2021a). Furthermore, the abundance of data may also bring about new methods for problem-solving, drive technological transformations, and lead the shift towards data-driven models, which are more aligned with the current state of our information society (Tang et al., 2021b). The widespread application of data has profoundly impacted our way of life, inevitably making "data" a primary driving force behind method evolution. Therefore, it is crucial for us to recognize the importance of data and proactively address the challenges and opportunities brought about by its abundance (Järvelin & Vakkari, 2021).

### *5.2 Implications*

We provide a quantitative analysis of the evolution of research methods related to data-driven using the field of LIS as a case study. The research design and methods employed in this study are universal, not only applicable for conducting similar studies on method evolution in other fields but also contributing to a deeper understanding of the development processes in different disciplines. Specifically, by focusing on fine-grained method entities related to data-driven, this study explores the evolution of research methods within the field from a more detailed perspective, offering a new viewpoint for this research field. In addition, this study uses research method entities in academic papers as the measurement unit. Through this measurement method and problem solving approach, it further refines the granularity and comprehensiveness of research methods within the field, providing a more comprehensive explanation for the evolution of research methods. The use of RMEs to illustrate specific evolutionary processes contributes to a more objective interpretation of the evolution of research methods in the current LIS field. In general, through the observation of fine-grained entities, we can better understand the evolutionary patterns of research methods in different academic fields, providing deeper insights and guidance for academic research.

### *5.3 Limitations*

This study utilized over 25,000 full-text articles from 14 academic journals in the LIS field as the research sample. Despite the large scale and extensive time span of the dataset compared to previous related studies, the corpus size used in this study is still insufficient when considering the theme of research method evolution in the field. Additionally, this study focused on discussing the evolutionary characteristics of research methods based on fine-grained method entities, categorizing these entities into algorithm and model, data resource, software and tool, and metric. This classification system of entities may not comprehensively describe the usage characteristics of research methods in the LIS field, which could result in the omission of fine-grained method entities in some academic papers.

## 6. Conclusion and future work

This study utilizes quantitative methods to analyze and summarize the evolutionary patterns of data-driven research methods in the LIS field from the perspective of fine-grained method entities. A corpus of full-text academic papers is constructed for this study, assuming that the methods mentioned in the Methods chapter of academic papers are more likely to be the research methods used in these papers. Therefore, it is necessary to identify the structural elements of academic papers and then use automated extraction methods to extract fine-grained method entities related to data-driven from the academic papers. Finally, the distribution of research method entities in the methods chapter is analyzed to examine the evolutionary characteristics of research methods. The research findings suggest that data resources are the primary driving force behind scholarly research, and the development of research methods in the LIS field exhibits a cyclical process of "emergence period-stable period/practice period".

In future research, it may be beneficial to explore the evolutionary characteristics of research methods using larger-scale data, which could potentially enhance the objectivity of the conclusions drawn in this study. By establishing a more universal and representative classification system for research method entities and enhancing the performance of entity extraction, it is possible to achieve a high adaptability to the usage characteristics of research methods in the described field. This may facilitate the discovery of additional evolutionary features of research methods based on the findings of this study.

## Acknowledgments

This work was supported by the National Natural Science Foundation of China (Grant No.72074113).

## Declarations

Conflict of interest: The author(s) declared no potential conflicts of interest with respect to the research, authorship, and/or publication of this article.

# Appendices

## Appendix A. Classification results of research methods

| Research method | #Article |
|---|---|
| Bibliometrics | 5750 |
| Content analysis | 3960 |
| Delphi study | 74 |
| Ethnography /field study | 236 |
| Experiment | 5081 |
| Focus group | 221 |
| Historical method | 917 |
| Interview | 1676 |
| Observation | 264 |
| Questionnaire | 3872 |
| Research diary / journal | 82 |
| Theoretical approach | 3946 |
| Think aloud protocol | 111 |
| Transaction log analysis | 419 |
| Webometrics | 712 |
| Others | 1074 |

## Appendix B. Statistical chart of full-text raw corpus chapter structure classification results

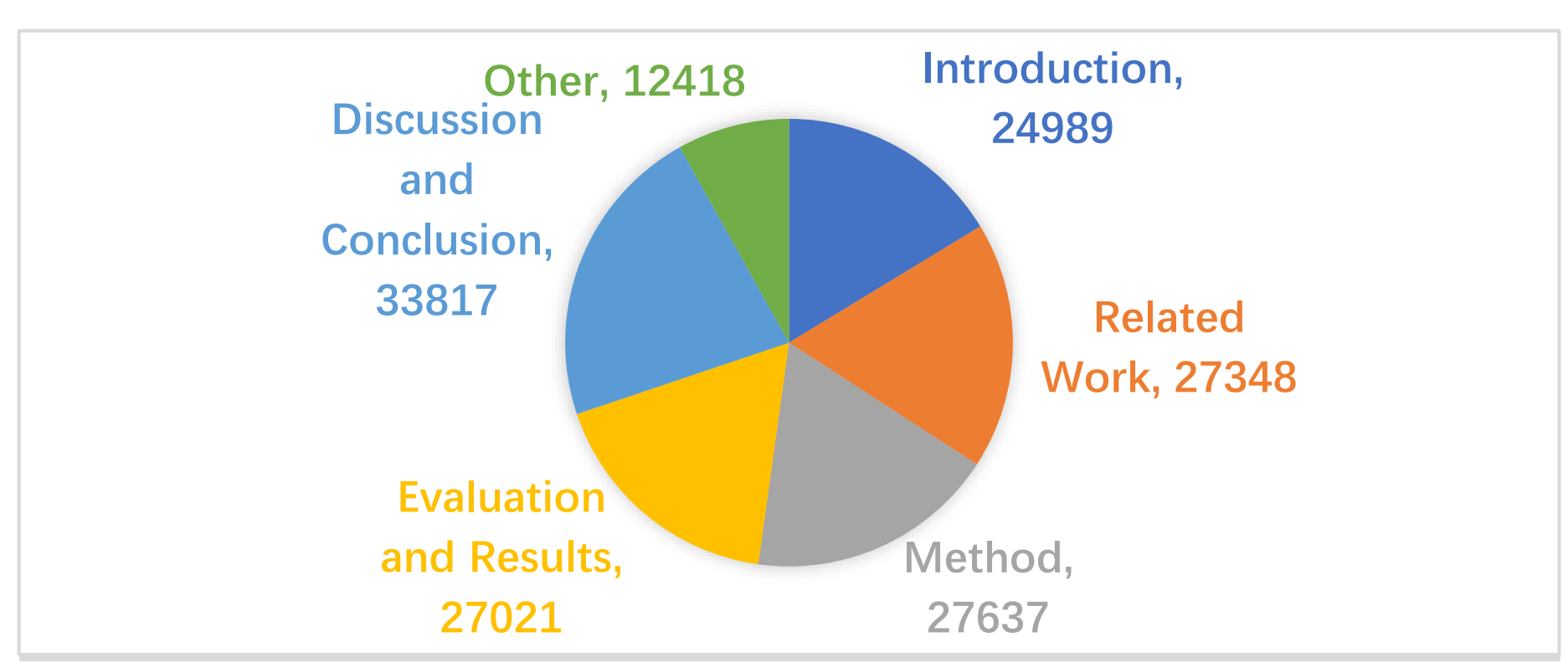

## Appendix C. Categories and definitions of fine-grained research method entities

| Type | Definition |
|---|---|
| Algorithm & Model | Nouns or noun phrases related to algorithms, models, designs, etc. mentioned in academic papers. |
| Data resource | Data, datasets, corpora, lexicons, dictionaries, or other resources mentioned in academic literature. |
| Software & Tool | Software, platforms, programming languages, algorithm packages, open-source tools, etc. mentioned in academic literature. |
| Metric | Evaluation-related metrics refer to the effectiveness of algorithms, models, designs, etc. mentioned in academic literature. |

## Appendix D. Number of fine-grained research method entity recognitions

| Type | # (RME terms) | # (Types of RME terms) |
|---|---|---|
| Algorithm & Model | 552282 | 90231 |
| Data resource | 1081612 | 232973 |
| Software & Tool | 547538 | 120379 |
| Metric | 799719 | 83558 |